\documentstyle[amscd,amssymb,verbatim,12pt]{amsart}
\def\dspace{\baselineskip=0.3 in}
\begin{document}
\dspace
\title[SCALE FACTOR DEPENDENT....]{SCALE FACTOR DEPENDENT EQUATION OF STATE
  FOR CURVATURE INSPIRED DARK ENERGY, PHANTOM BARRIER AND LATE COSMIC
  ACCELERATION }

\author{\bf S.K.Srivastava}
{ }
\maketitle
\centerline{ Department of Mathematics,}
 \centerline{ North Eastrn Hill University,}
 \centerline{  Shillong-793022, India}
\centerline{ srivastava@@.nehu.ac.in; sushil@@iucaa.ernet.in }

\vspace{1cm}

\centerline{\bf Abstract}

Here, it is found that dark energy and dark matter emerge from the
gravitational sector, if non-linear term of scalar curvature is added to
Einstein-Hilbert lagrangian. An equation of state for dark energy,
having the form $p_{\rm de} = - \rho_{\rm de} + f(a)$ (with $p_{\rm
  de}(\rho_{\rm de})$ being the pressure (density) for dark energy, $f(a)$
being a function of scale factor $a(t)$ and $t$ being the cosmic time) is
explored. Interestingly, this equation of state leads to a phantom barrier ${\rm w}_{\rm de} = p_{\rm
  de}/\rho_{\rm de} = - 1$  at $a = a_{\rm w}$. It is found that
when $a < a_{\rm w},{\rm w}_{\rm de} > -1 $ and ${\rm w}_{\rm de} < -1$ for $a
> a_{\rm w}$, showing a transition from non-phantom to phantom phase at $a =
a_{\rm w} < a_0 (a_0 $ being the current scale factor of the universe).


 \vspace{1.5cm}

Late cosmic accleration is the most remarkable astronomical observation
in the recent past \cite{sp,ag}. Theoretically, it is possible when the
universe is driven by a fluid having negative pressure and equation of state 
parameter (EOSP) ${\rm w}_{\rm de} < - 1/3.$ Observational data also indicate
that ${\rm w}_{\rm de}$ should be close to $-1$ (the phantom barrier) and most
probably less than $-1$ for the current universe. The source of this
mysterious fluid is still in dark. So, many phenomenological models, suggesting
possible sources of dark energy (DE), have been proposed in the past few years
\cite{vs,ms}. In spite of non-gravitational sources of DE, non-linear terms of
curvature are also proposed as gravitational alternative for DE \cite{snsd}. A
recent comprehensive review on DE is available in \cite{msej}.

In \cite{snsd}, non-linear terms of curvature are taken as DE lagrangian {\em
  a priori} and
its various consequences are discussed. In the present model,the story is
different, Here also, investigations begin from the modified gravity stemming
from addition of non-linear term to Einstein-Hilbert lagrangian, but unlike
\cite{snsd}, here, non-linear term of curvature is {\em not} taken as a DE source. Rather,
DE emerges spontaneously as a combined effect of linear as well as non-linear
terms of Ricci scalar curvature \cite{sks, skshep}, whereas, in \cite{snsd},
only non-linear term of curvature contributes to DE. Interestingly, here,it is found that
curvature can be a possible source of dark matter too. In \cite{sns}, various
 equations of state (EOS) for DE, dependent on Hubble's expansion rate $H$ and its
derivatives, are proposed. Contrary to this, in what follows, EOS for DE is
not proposed but derived. EOS for DE, obtained here, has the form $p_{\rm de} = - \rho_{\rm de}
+ f(a)$ ( with $\rho_{\rm de}, p_{\rm de}, a(t)$ and $t$ being dark energy
density, pressure, scale factor and cosmic time respectively). It is
interesting to see that $ a = a_{\rm w} < a_0$ ($a_0$ is the current scale
factor)  gives the phantom barrier ${\rm
  w}_{\rm de} = - 1$. It is found that when $a < a_{\rm w}, -1 < {\rm w}_{\rm
  de} < - 1/3 $ and ${\rm   w}_{\rm de} < - 1$ for $a > a_{\rm w}$. It shows a
{\em transition} from a quintessence to phantom phase at $a = a_{\rm w}$. In
\cite{sno}, a possibility of this type of transition is discussed taking EOS
in Jordan frame. The present work is different from \cite{sno} as
no EOS is taken here {\em a priori}, rather it is derived yielding phantom
  barrier and quintessence to phantom transition spontaneously.

Natural units ($\hbar = c = 1$) are used here with GeV as the fundamental
unit, where $\hbar$ and $c$ have their usual meaning. In this unit, it is
found that $1 {\rm GeV}^{-1} = 6.58 \times 10^{-25} {\rm sec}$.

\bigskip

\noindent 2.The action
for higher-derivative gravity is taken as 
$$ S_g = \int {d^4x} \sqrt{- g} \Big[ \frac{R}{16 \pi G} - \alpha R^{(2 + r)}
\Big], \eqno(1)$$
where $R$ is the Ricci scalar curvature, $G = M_P^{-2} ( M_P = 10^{19}$ GeV is
the Planck mass). Moreover,  $\alpha $ is a coupling constant having dimension
(mass)$^{- 2 r}$ with $r$ being a positive real number. As $(2 + r) > 0,$
instability problem does not arise like the model containing the $R^{-1}$
term \cite{add}. But, as mentioned above, here approach is different from the
papers \cite{sno,add}.  

The action (1) yields gravitational field equations
$$\frac{1}{16 \pi G} ( R_{\mu\nu} - \frac{1}{2} g_{\mu\nu} R ) - \alpha [ (2 +
r) \{
\triangledown_{\mu} \triangledown_{\nu}R^{(1 + r)} - g_{\mu\nu} {\Box} R^{(1 + r)} + R^{(1 + r)}
R_{\mu\nu} \}$$
$$- \frac{1}{2} g_{\mu\nu} R^{(2 + r)}] = 0 \eqno(2)$$
using the condition $\delta S_g/{\delta g^{\mu\nu}} = 0.$ Here, $\triangledown_{\mu}$
denotes covariant derivative and the operator $\Box$ is given as
$${\Box} = \frac{1}{\sqrt{-g}} \frac{\partial}{\partial x^{\mu}}
\Big(\sqrt{-g} g^{\mu\nu} \frac{\partial}{\partial x^{\nu}} \Big) \eqno(3)$$
with $\mu, \nu = 0,1,2,3$ and $g_{\mu\nu}$ as metric tensor components.

Taking trace of (2) and doing some manipulations, it is obtained that
$${\Box}R  + \frac{r}{R}\triangledown^{\nu}R \triangledown_{\nu}R = \frac{1}{3
  (2 + r)(1 + r)} \Big[\frac{R^{(1 - r)}}{16 \pi G \alpha} + r R^2 \Big]\eqno(4)$$
with $ \alpha \ne 0$ to avoid the ghost problem. 

Experimental evidences support spatially homogeneous flat model of the
universe \cite{ad}. So, the line-element, giving geometry of the universe, is
taken as
$$ dS^2 = dt^2 - a^2(t)[dx^2 + dy^2 + dz^2] \eqno(5)$$
with $a(t)$ as the scale factor. It gives expansion rate $H = {\dot a}/a.$

In the homogeneous space-time, given by (5), (4) is obtained as
$$ {\ddot R} + 3 \frac{\dot a}{a}{\dot R} + \frac{r {\dot R}^2}{R} = \frac{1}{3
  (2 + r)(1 + r)} \Big[\frac{R^{(1 - r)}}{16 \pi G \alpha} + r R^2 \Big]\eqno(6)$$

In most of the situations, for example, radiation model, matter-dominated
model, and accelerated models, we have $a(t)$ as a power-law solution yielding
$R$ as the power-law function of $a(t)$ . So, it is reasonable to take $R$
as
$$ R = \frac{A}{a^n}  \eqno(7)$$
with $n > 0$ being a real number and $A$ being a constant with mass
dimension 2.

$R$, given by (7), satisfies (6), if
$$ \frac{\ddot a}{a} + (2 - rn - n) \Big(\frac{\dot a}{a} \Big)^2 = \frac{1}{3
 (2 + r)(1 + r)} \Big[- \frac{a^{nr}}{8 \pi G n \alpha} - \frac{ r A}{n} a^{-n}\Big] \eqno(8)$$

Eq.(8) integrates to
$$ \Big(\frac{\dot a}{a} \Big)^2 = - \frac{C}{a^{2[3 - n (1 + r)]}} -  \frac{1}{3
  (2 + r)(1 + r)} \Big[- \frac{a^{nr}}{8 \pi G n \alpha ( nr + 2[3 - n (1 -
  r)])} A^{-r} $$
$$- \frac{ r A}{3 n(- n + 2[3 - n(1 + r)])} a^{-n}\Big] \eqno(9)$$
with $C$ being an integration constant. Setting $n = 3,$
(9) is re-written as
$$\Big(\frac{\dot a}{a} \Big)^2 = - C a^{6 r} + \frac{  A^r}{72 \pi G \alpha
  r(2 + r)(1 + r)} a^{3r} + \frac{r A}{27 (1 + 2 r) (2 + r) (1 + r)
  a^3}. \eqno(10)$$ 

This is the modified Friedmann equation giving dynamics of the universe. The
third term on r.h.s (the right hand side) of this equation has the form of density for
pressureless matter. This term emerges due to non-linear term of curvature in
the action (1), hence it is termed as dark matter density. So,
$$ \frac{8 \pi G}{3} \rho_{\rm dm} = \frac{r A}{27 (1 + 2 r) (2 + r) (1 + r)
  a^3}. \eqno(11)$$ 

The first term, on r.h.s. of (10), emerges spontaneously and  second term is
the combined effect of linear as well as non-linear term of curvature in the
action (1). A very interesting cosmic scenario is obtained on using these
two terms (first and second) and taking energy
density $\rho_{\rm de}$ as
$$\rho_{\rm de} = B - \frac{  A^r}{72 \pi G \alpha
  r(2 + r)(1 + r)} a^{3r} \Big[1 - \frac{a^{3r}}{2 \lambda} \Big] ,
  \eqno(12a)$$
with
$$ \lambda = \frac{A^r}{144 \pi G \alpha r(2 + r)(1 + r) C} . \eqno(12b)$$

If $\rho_{\rm de} = \rho_{\rm de(s)}$ at $a = a_s$ such that $a_s^{3r} = 2
\lambda$, (12a) looks like
$$ \rho_{\rm de} = \rho_{\rm de(s)} - \frac{  A^r}{72 \pi G \alpha
  r(2 + r)(1 + r)} a^{3r} \Big[1 - \frac{a^{3r}}{2 \lambda} \Big] ,
  \eqno(13)$$

According to WMAP \cite{abl}, current values of dark matter density and dark
energy density are $\rho^0_{\rm dm} = 0.23 \rho^0_{\rm cr}$ and $\rho^0_{\rm
  de} = 0.73 \rho^0_{\rm cr}$ with
$$ \rho^0_{\rm cr} = \frac{3 H_0^2}{8 \pi G},  \eqno(14)$$
$H_0 = 100h km/Mpc Sec = 2.32 \times 10^{-42}h {\rm GeV}$ and $h =
0.68$. Using these observational values in (11) and (13), we obtain
$$ \frac{8 \pi G}{3} \rho_{\rm dm} = 0.23 H_0^2 \Big(\frac{a_0}{a} \Big)^3
\eqno(15)$$ 
and
$$\rho_{\rm de} = \rho_{\rm de(s)} - (\rho_{\rm de(s)} - \rho^0_{\rm de}) \Big(\frac{a}{a_0} \Big)^{3r} \frac{\Big[1 - {a^{3r}}/{2 \lambda} \Big]}{\Big[1 - {a_0^{3r}}/{2 \lambda} \Big]} ,
  \eqno(16)$$
where
$$ \frac{  A^r}{72 \pi G \alpha r(2 + r)(1 + r)} = (\rho_{\rm de(s)} -
\rho^0_{\rm de}) \Big(a_0^{3r}\Big[1 - {a_0^{3r}}/{2 \lambda} \Big]
\Big)^{-1}. \eqno(17)$$ 

Using (11),(15) and (17), Friedmann equation (10) looks like
$$\Big(\frac{\dot a}{a} \Big)^2 = (\rho_{\rm de(s)} - \rho^0_{\rm de})
\Big(\frac{a}{a_0} \Big)^{3r} \frac{\Big[1 - {a^{3r}}/{2 \lambda}
  \Big]}{\Big[1 - {a_0^{3r}}/{2 \lambda} \Big]} + 0.23 H_0^2
\Big(\frac{a_0}{a} \Big)^3. \eqno(18)$$ 

If the term, proportional to $a^{-3}$, dominates other terms on the r.h.s. of
(18), it reduces to 
$$\Big(\frac{\dot a}{a} \Big)^2 \simeq  0.23 H_0^2
\Big(\frac{a_0}{a} \Big)^3 $$ 
yielding
$$ a(t) = a_d [1 + 0.72 H_0 a_d^{-3/2} (t - t_d) ]^{2/3} \eqno(19)$$
which shows decelerated expansion as ${\ddot a} < 0$. Here $a_d$ and $t_d$ are
constants.

When other terms dominate the term proportional to $a^{-3}$ on the r.h.s. of
(18) and $\rho_{\rm de(s)} > \rho^0_{\rm de}$, (18) reduces to
$$\Big(\frac{\dot a}{a} \Big)^2 = (\rho_{\rm de(s)} - \rho^0_{\rm de})
\Big(\frac{a}{a_0} \Big)^{3r} \frac{\Big[1 - {a^{3r}}/{2 \lambda}
  \Big]}{\Big[1 - {a_0^{3r}}/{2 \lambda} \Big]} . \eqno(20)$$ 
which is integrated to
$$ a(t) = a_* \Big[ \frac{a_*^{3 r}}{2 \lambda} + \Big\{\sqrt{1 - \frac{a_*^{3
      r}}{2 \lambda}} - \frac{3}{2} r B a_0^{3r/2} (t - t_0) \Big\}^2 \Big]^{-
      1/3r} \eqno(21)$$
giving acceleration as ${\ddot a} > 0$. Here, $t_*$ is the time for {\em
      transition from deceleration to acceleration} and $a_*$ is the
      corresponding scale factor. Moreover, the constant $B$ in (21) is given
      by 
$$ B^2 = (\rho_{\rm de(s)} -
\rho^0_{\rm de}) \Big(a_0^{3r}\Big[1 - {a_0^{3r}}/{2 \lambda} \Big]
\Big)^{-1}. \eqno(22)$$ 

It is obvious that for $t > t_*$, (18) reduces to (20). So, if $\rho_{\rm
  de(s)} < \rho^0_{\rm de}$, (20) shows that $\Big(\frac{\dot a}{a} \Big)^2 < 0$
  for $t > t_*$. It leads to an {\em unphysical} situation. So, this
  possibility is rejected and it is concluded that
$$\rho_{\rm de(s)} >  \rho^0_{\rm de} . \eqno(23)$$

Further, DE conservation equation is given as
$$ {\dot \rho_{\rm de}} + 3 H (\rho_{\rm de} + p_{\rm de} ) = 0 . \eqno(24)$$

Connecting (16) and (24), it is obtained that
$$p_{\rm de} = - \rho_{\rm de} + (\rho_{\rm de(s)} - \rho^0_{\rm de}) \Big(\frac{a}{a_0} \Big)^{3r} \frac{\Big[1 - {a^{3r}}/{ \lambda} \Big]}{\Big[1 - {a_0^{3r}}/{2 \lambda} \Big]} ,
  \eqno(25)$$
This equation shows that at $a = \lambda^{1/3r} = a_{\rm w}, p_{\rm de} = -
  \rho_{\rm de}.$ So,
$$p_{\rm de} = - \rho_{\rm de} + (\rho_{\rm de(s)} - \rho^0_{\rm de}) \Big(\frac{a}{a_0} \Big)^{3r} \frac{\Big[1 - \Big({a}/{a_{\rm w} }\Big)^{3r} \Big]}{\Big[1 - \frac{1}{2}\Big({a_0}/{a_{\rm w} }\Big)^{3r} \Big]} ,
  \eqno(26)$$
This is the EOS for DE, obtained in this model, which is scale factor dependent.
  $a_s$, given above,  and $a_{\rm w}$ are related as $ a_s = 2^{1/3r} a_{\rm
  w} > a_0$ and $ a_* < a_{\rm w} < a_0.$

Now, (16) and (26) yield
$$\rho_{\rm de} + 3 p_{\rm de} = - 2 \rho_{\rm de(s)} +  (\rho_{\rm de(s)} -
  \rho^0_{\rm de}) \Big(\frac{a}{a_0} \Big)^{3r} \frac{\Big[1 - 2 \Big({a}/{a_{\rm w} }\Big)^{3r} \Big]}{\Big[1 - \frac{1}{2}\Big({a_0}/{a_{\rm w} }\Big)^{3r} \Big]} ,
  \eqno(27)$$ 

Due to inequality (23), (26) and (27) yield
$$\rho_{\rm de} +  p_{\rm de} > 0 , \eqno(28a)$$
$$\rho_{\rm de} +  3 p_{\rm de} < 0 , \eqno(28b)$$
for $a_* < a <  a_{\rm w}$ and
$$\rho_{\rm de} +  p_{\rm de} < 0 , \eqno(29a)$$
$$\rho_{\rm de} +  3 p_{\rm de} < 0 , \eqno(29b)$$
for $ a >  a_{\rm w}$.

(28a,b) give {\em quintessence} phase of DE, whereas (29a,b) give {\em
  phantomce} phase of DE. These two phases are divided by
$$  p_{\rm de} = - \rho_{\rm de} \eqno(30)$$
at $a = a_{\rm w}$ as  given by (27). Thus, we have transition from {\em
  non-phantom} to {\em phantom} at $a = a_{\rm w} > a_0$ 

It is given above that at $a_s^{3r} = 2 \lambda = 2 a^{3r}_{\rm w}.$ So, (18)
is obtained as
$$\Big(\frac{\dot a}{a} \Big)^2 = (\rho_{\rm de(s)} -
  \rho^0_{\rm de}) \Big(\frac{a}{a_0} \Big)^{3r} \frac{\Big[1 - 2
  \Big({a}/{a_{\rm s} }\Big)^{3r} \Big]}{\Big[1 - \Big({a_0}/{a_{\rm s}
  }\Big)^{3r} \Big]}    + 0.23 H_0^2 \Big(\frac{a_0}{a} \Big)^3. \eqno(31)$$  
(31) shows that when $a(t)$ reaches $a_{\rm s}$, acceleration given by (20)
  stops and deceleration driven by matter resumes. It shows {\em transient}
  acceleration as obtained in \cite{skshep}. Sahni \cite{vsaa} had obtained
  this type of result in the context of brane-gravity cosmology earlier. 

(31) also shows that $ {\dot a} = 0$ at $ a = a_{\rm m} > a_{\rm s}$ if
$$(\rho_{\rm de(s)} -
  \rho^0_{\rm de}) \Big(\frac{a_{\rm m}}{a_0} \Big)^{3r} \frac{\Big[1 - 2
  \Big({a_{\rm m}}/{a_{\rm s} }\Big)^{3r} \Big]}{\Big[1 - \Big({a_0}/{a_{\rm s}
  }\Big)^{3r} \Big]}    + 0.23 H_0^2 \Big(\frac{a_0}{a_{\rm m}} \Big)^3 = 0
  . \eqno(32)$$
It shows that expansion will reach its maximum at $a = a_{\rm m}$ and it will
  begun to contract taking a turn around.

Thus, a transition from {\em deceleration} to {\em   acceleration}, at some
time $t_*$ in the recent past, is obtained giving a possible explanation for
late cosmic acceleration \cite{ag}. EOS for DE, derived here, depends on the
scale factor $a(t)$. Interestingly, it is found that another transition, from
  quintessence phase of DE to phantom phase, takes place at $a = a_{\rm w}$
  with $  a_* < a_{\rm w} < a_0$. Characteristics of quintessence DE are
  different from phantom one. Kinetic energy for the latter is negative,
  whereas it is positive for the former. Moreover, ${\rm w}_{\rm de} > - 1$
  for the former, ${\rm w}_{\rm de} < - 1$ for the latter. So, it indicates
  that possibly, DE has two components (i) quintessence and (ii)
  phantom. Former dominates when $a_* < a < a_{\rm w}$ and latter dominates
  when $ a_{\rm w} < a < a_{\rm s}$. It is found that, after acceleration for
  some time, universe decelerates . The decelerated expansion continues till
  scale factor acquires its maximum at time $t_m$ and will begin to contract
  for $t > t_m$.

\end{document}